\documentclass[11pt,english,aps]{revtex4}
\usepackage{times}
\usepackage[T1]{fontenc}
\usepackage[latin1]{inputenc}
\usepackage{subfigure}
\usepackage{graphicx}

\makeatletter

\providecommand{\tabularnewline}{\\}

\usepackage{babel}
\makeatother
\begin{document}

\title{A compound figure of merit for photonic applications of metal nanocomposites}

\author{Hernando Garcia%
\thanks{Contact author at hgarcia@siue.edu%
}}

\address{Department of Physics Southern Illinois University Edwardsville, IL 62026 }

\author{Hare Krishna and Ramki Kalyanaraman }

\address{Department of Physics Washington University in St. Louis St. Louis, MO 63130 }

\begin{abstract}
Selecting nanocomposites for photonic switching applications requires optimizing their thermal,
nonlinear and two-photon absorption characteristics. We simplify this step by defining a compound
figure of merit ($FOM_{C}$) for nanocomposites of noble metals in dielectric based on criteria
that limit these structures in photonic applications, i.e. thermal heating and two-photon absorption.
The device independent results predict extremely large values of $FOM_{C}$ for a specific combination
of the metal and insulator dielectric constant given by $\epsilon_{h}=\frac{\epsilon_{1}-\epsilon_{2}}{2}$,
where $\epsilon_{h}$ is the dielectric constant of the host and $\epsilon_{1}$ and $\epsilon_{2}$
are the real and imaginary parts for the metal. \pagebreak
\thispagestyle{empty}
\end{abstract}
\maketitle
In recent years considerable research has gone into the study of all-optical switching devices
for photonic applications where the requirement of the material had been a large $n_{2}$ (nonlinear
refractive index), low index change due to thermal effects, and small two-photon absorption (TPA)
\cite{friberg87,stegeman88}. Furthermore, there is significant interest in using nanocomposites
comprising of metal nanoparticles in dielectric host for such applications because of their nonlinear
response \cite{yang93,yang96,wang03} and because of their potential use in applications below
the diffraction limited regime \cite{haglund88,maier03,huter04}. However, the selection of the
metal-host system for a given application must be based on a careful evaluation of the key physical
limitations in the interaction of light with matter. In this regard, there are two well accepted
limitations. The first is the refractive index change due to thermal effects arising from light-matter
interactions \cite{friberg87}. The second, which has been recognized as the ultimate limit for
optical switching, is TPA \cite{mizrahi89}. However, there has been no effort to combine these
effects into a figure of merit describing the metal nanocomposite material in photonic applications
involving optical modulation \cite{hartfield78}. 

In this letter we define a compound figure of merit ($FOM_{C}$) as the product of the thermal
FOM ($FOM_{Therm}$) and the TPA FOM ($FOM_{TPA}$). The reason for choosing the product as compared
to a linear combination comes from the strong correlation between the TPA and thermal effects.
In general material having large TPA show small linear absorption and thermal effects and vice
versa. Therefore it is difficult to select materials for such applications based simply on the
linear combination of the two effects. A common example of such a definition is that of the gain-bandwidth
product used to optimize an op-amp. the The theoretical behavior for Ag, Au and Cu nanoparticles
embedded in an SiO$_{\textrm{2}}$ matrix were evaluated based on this $FOM$. The behavior of
the $FOM_{C}$, which is device independent, shows that an extremely large value results at a
specific combinations of the metal and host dielectric function given by $\epsilon_{h}=\frac{\epsilon_{1}-\epsilon_{2}}{2}$,
where $\epsilon_{h}$ is the dielectric constant of the host and $\epsilon_{1}$ and $\epsilon_{2}$
are the real and imaginary parts of the the dielectric constant of the metal. These results could
simplify the process of materials selection for nanocomposites in photonic switching applications
at desired wavelengths. 

We start with the criteria that conventional all-optical switching devices are based on generation
of light induced phase shifts of the order of $p\pi$, where $4\geq p\geq1$ for a range of devices.
The design in mind for these calculations is a Mach-Zehnder interferometer where $p=1$, and
the criterion for switching is given by \cite{maier03}:\begin{equation}
p\pi=\frac{2\pi n_{2}I_{o}L}{\lambda}\label{eq:ppi}\end{equation}

where $I_{o}$ is the switching intensity, $L$ is the length of the device, and $\lambda$ is
the wavelength. Our first figure of merit requires that the thermal index change be a small fraction
of the fast light-induced change ($n_{2}I_{o}$) \cite{friberg87}. The thermal index change
can be expressed by:\begin{equation}
thermal\, index\, change=\frac{\partial n}{\partial T}\frac{\Delta Q}{C_{p}\rho V}\label{eq:thermalindexchange}\end{equation}
 where $\frac{\partial n}{\partial T}$ is the thermo-optic coefficient, $C_{p}$ is the specific
heat, $V$ is the volume of the material heated by the light, $\rho$ is the density of the material
and $\Delta Q$ is the energy absorbed by the material per pulse, which is given by: \begin{equation}
\Delta Q=\alpha LI_{o}A\tau\label{eq:deltaQ}\end{equation}

where $A$ is the cross sectional area of the device, $\alpha$ is the linear absorption coefficient
and $\tau$ is the duration of the switching pulse. By using eq. \ref{eq:thermalindexchange}
and \ref{eq:deltaQ} the $FOM_{Therm}$ becomes: \begin{equation}
FOM_{Therm}=\frac{n_{2}C_{p}\rho}{\frac{\partial n}{\partial T}\tau\alpha}\label{eq:FOMtherm}\end{equation}

The behavior of this FOM for Au, Ag and Cu nanoparticles embedded in SiO$_{\textrm{2}}$ is shown
in Fig. \ref{cap:Figureofmerit}(a). The values of various parameters are listed in tables \ref{cap:Optical-properties}
and \ref{tab:MetalProperties}. 

Two-photon absorption can reduce the device throughput even if the linear absorption is small.
A general criterion for this was developed by Stegeman and co-workers \cite{stegeman88} based
on Eq. \ref{eq:ppi} and the two photon absorption coefficient $\beta$ as:\[
\frac{\beta\lambda}{2n_{2}}<1\, or\,\frac{2n_{2}}{\beta\lambda}>1\]

Based on this, we defined our next figure of merit to be:\begin{equation}
FOM_{TPA}=\frac{2n_{2}}{\beta\lambda}\label{eq:FOMTPA}\end{equation}
 The behavior of this quantity is shown in Fig. \ref{cap:Figureofmerit}(b) \emph{}for the noble
metal nanoparticles. 

The compound figure of merit in this letter takes into account the switching limitations due
to thermal heating and large two photon absorption and is defined as the product of Eq. \ref{eq:FOMtherm}
and Eq. \ref{eq:FOMTPA} to give: \begin{equation}
FOM_{C}=FOM_{Therm}.FOM_{TPA}=\frac{2n_{2}^{2}C_{p}\rho}{\alpha\tau\beta\lambda\frac{\partial n}{\partial T}}\label{eq:FOMC}\end{equation}
 \emph{Using this definition, an attractive combination of metal-dielectric would be one with
a large value of the $FOM_{C}$}. It is natural to introduce this figure of merit because there
could be materials with excellent thermal properties but very large $TPA$, or vice versa, or
various other combinations, including both high or both low. On the other hand materials showing
a trade off between moderate $TPA$ and moderate thermal heating may be excellent candidates
for photonic applications where the only limitations will be device design and optimization. 

Using this new $FOM_{C}$, we evaluated the response for metal nanocomposites. In these structures
we have a lot of freedom in choosing and controlling $\alpha$, $\beta$, and $n_{2}$, which
are the only free parameters in the above figure of merit. In the evaluation we assumed the following: 

\begin{enumerate}
\item $\rho$, $C_{p}$, and $\frac{\partial n}{\partial T}$ can be considered to be properties of
the host only and are not influenced by the presence of the metal nanostructures (i.e. \emph{dilute
limit approximation}); 
\item $\beta$, $\alpha$, and $n_{2}$ are drastically modified by dielectric confinement, and they
may be analyzed within the context of Mie theory \cite{born80}; 
\item The particle diameters are such that quantum confinement effects due to intra-band dipole electronic
transitions are negligible; 
\item The main contribution to the nonlinear susceptibility comes from the inter-band electric dipole
transition, and the contribution is negative, and purely imaginary. 
\end{enumerate}
With the above assumptions we calculated $FOM_{C}$ (Eq. \ref{eq:FOMC}) for a metal nanocomposite,
by noting the following: \begin{equation}
\frac{n_{2}}{\beta\lambda}=\frac{Re[\chi_{eff}^{(3)}]}{4\pi Im[\chi_{eff}^{(3)}]}\label{eq:n2overbetalambda}\end{equation}

where we have ignored, for the moment, the frequency dependence of the nonlinear susceptibilities.
$\chi_{eff}^{(3)}$ is the effective nonlinear susceptibility and is given in the context of
Mie's theory by:\begin{equation}
\chi_{eff}^{(3)}=vf_{1}^{2}\mid f_{1}\mid^{2}\chi_{m}^{(3)}\label{eq:Chieff}\end{equation}
 where $v$ is the volume fraction of the metal in the host matrix, $\chi_{m}^{(3)}$ is the
third order susceptibility of the metal cluster, and $f_{1}$ is an enhancement factor produced
by the dielectric confinement, which can be expressed as: \begin{equation}
f_{1}=\frac{3\epsilon_{h}}{\epsilon_{1}+i\epsilon_{2}+2\epsilon_{h}}\label{eq:f1}\end{equation}
 where $\epsilon_{h}$ is the real dielectric constant of the host, and $\epsilon_{1}$ and $\epsilon_{2}$
are the real and imaginary dielectric constants of the metal nanoparticles respectively. Using
Eq. \ref{eq:Chieff} in Eq. \ref{eq:n2overbetalambda} and using assumption (4) we get for the
ratio of the real and imaginary parts in Eq. \ref{eq:n2overbetalambda}: \begin{equation}
\frac{Re[\chi_{eff}^{(3)}]}{Im[\chi_{eff}^{(3)}]}=-\frac{Im[f_{1}^{2}]}{Re[f_{1}^{2}]}\label{eq:ReoverImChi}\end{equation}
 We calculated the ratio of $n_{2}$ \cite{sutherland96} and $\alpha$ \cite{yang96}, using
the following definitions: \begin{equation}
\alpha=\frac{v\omega\mid f_{1}\mid^{2}\epsilon_{2}}{n_{o}c}\label{eq:alpha}\end{equation}
 and\begin{equation}
n_{2}=\frac{12\pi}{n_{o}^{2}c}Re[\chi_{eff}^{(3)}]=\frac{12\pi}{n_{o}^{2}c}v\mid f_{1}\mid^{2}Re[f_{1}^{2}\chi_{m}^{(3)}]\label{eq:n2}\end{equation}
 where, $n_{o}$ is the refractive index, $c$ is the velocity of light and $\omega$ is the
frequency of the incident light. This results in a ratio of:\begin{equation}
\frac{n_{2}}{\alpha}=\frac{6\lambda\mid\chi_{m}^{(3)}\mid}{n_{o}c\epsilon_{2}}Re[if_{1}^{2}]\label{n2overalpha}\end{equation}
 By using the above result and Eq. \ref{eq:f1} in Eq. \ref{eq:FOMC} we get the compound FOM
as: \begin{equation}
FOM_{C}=\frac{108C_{p}\rho\mid\chi_{m}^{2}\mid}{\pi\tau\frac{\partial n}{\partial T}n_{o}c}\left\{ \frac{\lambda\epsilon_{2}\epsilon_{h}^{2}(\epsilon_{1}+2\epsilon_{h})^{2}}{\left[(\epsilon_{1}+2\epsilon_{h})^{2}-\epsilon_{2}^{2}\right]\left[(\epsilon_{1}+2\epsilon_{h})^{2}+\epsilon_{2}^{2}\right]^{2}}\right\} \label{FOMCfinal}\end{equation}
 The dependence of the $FOM_{C}$ on the dielectric constants has been put in the brackets. As
would be expected, this $FOM_{C}$ goes to zero under the highly absorbing conditions found at
the plasmon resonance condition of $\epsilon_{1}+2\epsilon_{h}=0$. On the other hand, as Fig.
\ref{cap:Figureofmerit}(c), large maxima are observed in the wavelength range of $0.3$ to $0.6\,\mu m$
for the various metals using the parameters from table \ref{cap:Optical-properties}. The large
maxima occur at the condition of: \[
(\epsilon_{1}+2\epsilon_{h})^{2}-\epsilon_{2}^{2}=0\]
 The wavelengths corresponding to these maxima for each metal are tabulated in table \ref{cap:Location-of-peaks},
and are compared to nearby peak positions in the $FOM_{Therm}$ and $FOM_{TPA}$. A significant
difference in wavelength of the $FOM_{C}$ to those in the thermal or TPA cases is observed.
This difference is especially important when put in the context of data communication bandwidths,
where wavelength separations of 1 to 5 nm are routine. The physical origin of these wavelengths
can be understood by analyzing the location of the peaks within the framework of the simplifying
Drude model. In this model the real and imaginary components of the dielectric susceptibility
of the metal in the high frequency limit ($\omega\tau>>1$) are given by: \[
\epsilon_{1}=\epsilon_{o}-\frac{\omega_{p}^{2}}{\omega^{2}}\,\, and\,\,\epsilon_{2}=\frac{\omega_{p}^{2}}{\omega^{2}\tau}\]
where $\epsilon_{o}$ is the contribution to the dielectric constant of the metal from bound
charges and $\omega_{p}$ is the bulk plasmon resonance of the metal$.$ Using the condition
for the $FOM_{C}$ maxima and the condition for surface plasmon resonance given by:\[
\frac{\omega^{2}}{\omega_{sp}^{2}}=\epsilon_{1}+2\epsilon_{h}\]

After some algebraic manipulation, the frequency $\omega_{FOM}$ at which the maxima in the $FOM_{C}$
occurs can be related to the surface plasmon frequency $\omega_{sp}$ by the expression: \begin{equation}
(\frac{1}{\omega_{sp}^{2}}-\frac{1}{\omega_{FOM}^{2}})^{2}=(\frac{1}{\omega_{FOM}^{3}\tau})^{2}\label{eq:omegaFOM}\end{equation}

In eq. \ref{eq:omegaFOM}, the quantity on the right side is always positive, but the quantity
on the left side can be positive or negative giving: \begin{equation}
\frac{\omega_{FOM}^{2}}{\omega_{sp}^{2}}=(1\pm\frac{1}{\omega_{FOM}\tau})\label{eq:omegaFOM2}\end{equation}

Therefore we find that the maxima in the $FOM_{C}$ can be located above or below the surface
plasmon resonance. In the evaluation of eq. \ref{FOMCfinal} we have chosen the positive values
obtained from the above expression. In table \ref{cap:Location-of-peaks} we have also indicated
$\lambda{}_{FOM}^{Drude}$ estimated using the above Drude model for the various metals using
experimental values for $\lambda_{sp}$ and the relaxation times from Ref. \cite{johnson72}.
It is apparent that this crude model also predicts a shift in the peak position from $\lambda_{sp}$,
$\lambda_{Therm}$ and $\lambda_{TPA}$ and actually predicts very well the $\lambda_{FOM}$
position for the case of Cu estimated from Eq. \ref{FOMCfinal}. 

Recently there have been reports of lasing of Au nanoparticles in water \cite{kang06}. The authors
estimated a surface plasmon resonance for their Au-water system to be at 700 nm while the observed
lasing occurred with various peaks located at 720, nm or higher. Using our model, we have estimated
that the $FOM_{C}$ achieves a maximum for Au nanoparticles in water at $720\, nm$ for an electron
relaxation time of $9.3\times10^{-5}s$. While this may be rather coincidental, it should be
emphasized that the $FOM_{C}$ expressed by eq. \ref{FOMCfinal} represents a situation where
losses are minimized and the nanoparticles act as excellent scattering centers. 

In conclusion, we have defined a compound figure of merit $FOM_{C}$ for metal-dielectric nanocomposites
that takes into account limitations to optical switching from thermal index changes and two photon
absorption. As expected the $FOM_{C}$ is zero at the plasmon resonance when large absorption
occurs but has extremely large values at a value of frequency which is dependent only upon the
dielectric constants of the metal and dielectric. This result could allow the rapid selection
of a metal-host system for operation at a desired wavelength.

One of the authors (RK) acknowledges support by the National Science Foundation through grant
\# DMI-0449258.\pagebreak

\pagebreak

\pagestyle{empty}

\begin{table}[!tph]

\caption{Substrate  properties used in calculation of the $FOM$. \label{cap:Optical-properties}}

\begin{centering}\begin{tabular}{|c|c|}
\hline 
Substrate Parameters&
Value\tabularnewline
\hline
\hline 
Specific heat capacity, $C_{p}$ ($\times10^{6}ergs/g-K$)&
$7.4$\tabularnewline
\hline 
Density, $\rho$ ($g-cm^{-3}$)&
2.33 \tabularnewline
\hline 
Thermo-optic coefficient,

$\partial n/\partial T$ ($\times10^{-6}K^{-1}$)&
$0.55$\tabularnewline
\hline 
Refractive index, $n_{o}$&
1.46\tabularnewline
\hline 
Host dielectric constant, $\epsilon_{h}$&
3.82\tabularnewline
\hline
\end{tabular}\par\end{centering}
\end{table}

\begin{table}[!tph]

\caption{Materials properties for the metals used in the calculation of the $FOM$. \label{tab:MetalProperties}}

\begin{centering}\begin{tabular}{|c|c|c|c|}
\hline 
Metal Parameters&
Ag&
Au&
Cu\tabularnewline
\hline
\hline 
Fermi velocity, $v_{F}$ ($\times10^{8}m/s$)&
$1.42$&
$1.4$&
$1.28$\tabularnewline
\hline 
Electron density, $\rho_{e}$ ($\times10^{22}\#/cm^{3}$) &
$5.85$$ $&
$5.9$&
$8.45$\tabularnewline
\hline 
Effective mass, $m_{eff}$&
$0.96m_{o}$&
$0.99m_{o}$&
$0.99m_{o}$\tabularnewline
\hline 
High frequency dielectric constant, $\epsilon$&
5.3&
8.6&
9.3\tabularnewline
\hline 
Radius \& interparticle spacing (nm)&
35, 75&
35, 75&
35, 75\tabularnewline
\hline
\end{tabular}\par\end{centering}
\end{table}

\begin{table}[!tph]

\caption{Values of the wavelengths corresponding to the surface plasmon resonance ($\lambda_{sp}$) and
strongest peaks in the $FOM_{Therm}$ ($\lambda_{Therm}$), $FOM_{TPA}$ ($\lambda_{TPA}$) and
$FOM_{C}$ ($\lambda_{C}$). The position of the strongest peak as predicted by the crude Drude
model is expressed as $\lambda_{FOM}^{Drude}$. \label{cap:Location-of-peaks}}

\begin{centering}\begin{tabular}{|c|c|c|c|c|c|}
\hline 
Metal&
$\lambda_{sp}\,(nm)$&
$\lambda_{Therm}\,(nm)$&
$\lambda_{TPA}\,(nm)$&
$\lambda_{C}\,(nm)$&
$\lambda_{FOM}^{Drude}\,(nm)$\tabularnewline
\hline
\hline 
Ag&
487&
450&
472&
509&
497\tabularnewline
\hline 
Au&
552&
583&
550&
591&
560\tabularnewline
\hline 
Cu&
591&
618&
618&
607&
606\tabularnewline
\hline
\end{tabular}\par\end{centering}
\end{table}

\pagebreak

\pagestyle{empty}

\section*{Figure Captions}

\begin{figure}[!tph]
\begin{centering}\subfigure[]{\includegraphics[height=2in,keepaspectratio]{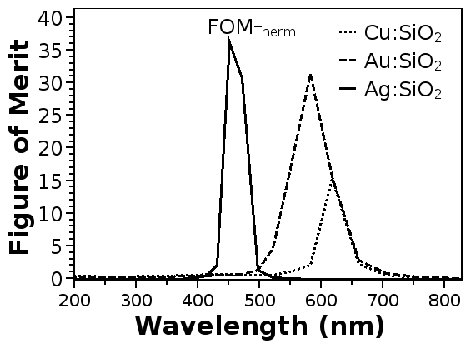}}\par\end{centering}

\begin{centering}\subfigure[]{\includegraphics[height=2in,keepaspectratio]{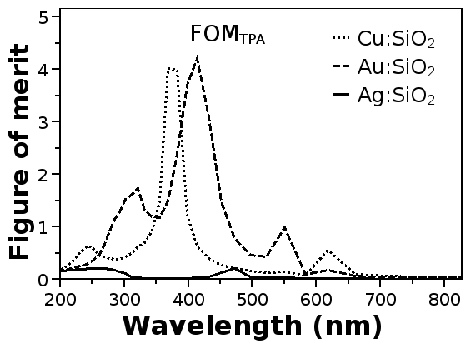}}\par\end{centering}

\begin{centering}\subfigure[]{\includegraphics[height=2in,keepaspectratio]{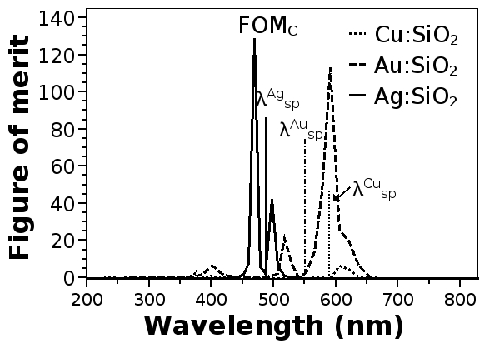}}\par\end{centering}

\caption{\label{cap:Figureofmerit}}
\end{figure}

\begin{itemize}
\item Fig. \ref{cap:Figureofmerit}: (a) The thermal FOM, $FOM_{Therm}$ for Ag (solid line), Au (dashed
line) and Cu (dotted line) embedded in SiO$_{\textrm{2}}$ matrix as a function of wavelength.
The values for the complex dielectric constant were obtained from ref. \cite{johnson72}. The
$FOM_{Therm}$ for Ag (solid line) has been scaled by a factor of $1/50$. (b) The two-photon
absorption FOM, $FOM_{TPA}$ for Ag, Au and Cu in SiO$_{\textrm{2}}$ as a function of wavelength.
(c) Total figure of merit given by Eq. \ref{FOMCfinal} for Ag (solid line), Au (dashed line)
and Cu (dotted line) embedded in SiO$_{\textrm{2}}$ matrix. The wavelengths corresponding to
the surface plasmon wavelengths are also indicated for each metal. 
\end{itemize}
\pagebreak

\pagebreak

\pagestyle{empty}

\begin{thebibliography}{14}
\expandafter\ifx\csname natexlab\endcsname\relax\def\natexlab#1{#1}\fi
\expandafter\ifx\csname bibnamefont\endcsname\relax
  \def\bibnamefont#1{#1}\fi
\expandafter\ifx\csname bibfnamefont\endcsname\relax
  \def\bibfnamefont#1{#1}\fi
\expandafter\ifx\csname citenamefont\endcsname\relax
  \def\citenamefont#1{#1}\fi
\expandafter\ifx\csname url\endcsname\relax
  \def\url#1{\texttt{#1}}\fi
\expandafter\ifx\csname urlprefix\endcsname\relax\def\urlprefix{URL }\fi
\providecommand{\bibinfo}[2]{#2}
\providecommand{\eprint}[2][]{\url{#2}}

\bibitem[{\citenamefont{Friberg and Smith}(1987)}]{friberg87}
\bibinfo{author}{\bibfnamefont{S.~R.} \bibnamefont{Friberg}} \bibnamefont{and}
  \bibinfo{author}{\bibfnamefont{P.~W.} \bibnamefont{Smith}},
  \bibinfo{journal}{IEEE J Quant. Elect.} \textbf{\bibinfo{volume}{23}},
  \bibinfo{pages}{2089} (\bibinfo{year}{1987}).

\bibitem[{\citenamefont{Stegeman et~al.}(1988)\citenamefont{Stegeman, Wright,
  Finlayson, Zanonic, and Seaton}}]{stegeman88}
\bibinfo{author}{\bibfnamefont{G.~I.} \bibnamefont{Stegeman}},
  \bibinfo{author}{\bibfnamefont{E.~M.} \bibnamefont{Wright}},
  \bibinfo{author}{\bibfnamefont{N.}~\bibnamefont{Finlayson}},
  \bibinfo{author}{\bibfnamefont{R.}~\bibnamefont{Zanonic}}, \bibnamefont{and}
  \bibinfo{author}{\bibfnamefont{C.~T.} \bibnamefont{Seaton}},
  \bibinfo{journal}{J. Lightwave. Tech.} \textbf{\bibinfo{volume}{6}},
  \bibinfo{pages}{953} (\bibinfo{year}{1988}).

\bibitem[{\citenamefont{Yang}(1993)}]{yang93}
\bibinfo{author}{\bibfnamefont{L.}~\bibnamefont{Yang}}, Ph.D. thesis,
  \bibinfo{school}{Vanderbilt University} (\bibinfo{year}{1993}).

\bibitem[{\citenamefont{Yang et~al.}(1996)\citenamefont{Yang, Osborne,
  Haglund~Jr, Magruder, White, Zuhr, and Hosono}}]{yang96}
\bibinfo{author}{\bibfnamefont{L.}~\bibnamefont{Yang}},
  \bibinfo{author}{\bibfnamefont{D.~H.} \bibnamefont{Osborne}},
  \bibinfo{author}{\bibfnamefont{R.~F.} \bibnamefont{Haglund~Jr}},
  \bibinfo{author}{\bibfnamefont{R.~H.} \bibnamefont{Magruder}},
  \bibinfo{author}{\bibfnamefont{C.~W.} \bibnamefont{White}},
  \bibinfo{author}{\bibfnamefont{R.~A.} \bibnamefont{Zuhr}}, \bibnamefont{and}
  \bibinfo{author}{\bibfnamefont{H.}~\bibnamefont{Hosono}},
  \bibinfo{journal}{Appl. Phys. A} \textbf{\bibinfo{volume}{62}},
  \bibinfo{pages}{403} (\bibinfo{year}{1996}).

\bibitem[{\citenamefont{Wang et~al.}(2003)\citenamefont{Wang, Chen, Yang, Guan,
  Yang, Zhou, and Lu}}]{wang03}
\bibinfo{author}{\bibfnamefont{W.~T.} \bibnamefont{Wang}},
  \bibinfo{author}{\bibfnamefont{Z.~H.} \bibnamefont{Chen}},
  \bibinfo{author}{\bibfnamefont{G.}~\bibnamefont{Yang}},
  \bibinfo{author}{\bibfnamefont{D.~Y.} \bibnamefont{Guan}},
  \bibinfo{author}{\bibfnamefont{G.~Z.} \bibnamefont{Yang}},
  \bibinfo{author}{\bibfnamefont{Y.~L.} \bibnamefont{Zhou}}, \bibnamefont{and}
  \bibinfo{author}{\bibfnamefont{H.~B.} \bibnamefont{Lu}},
  \bibinfo{journal}{Applied Physics Letters} \textbf{\bibinfo{volume}{83}},
  \bibinfo{pages}{1983} (\bibinfo{year}{2003}).

\bibitem[{\citenamefont{Haglund~Jr}(1988)}]{haglund88}
\bibinfo{author}{\bibfnamefont{R.~F.} \bibnamefont{Haglund~Jr}},
  \bibinfo{journal}{Mat. Sci. Eng.} \textbf{\bibinfo{volume}{253}},
  \bibinfo{pages}{275} (\bibinfo{year}{1988}).

\bibitem[{\citenamefont{Maier et~al.}(2003)\citenamefont{Maier, Kik, Atwater,
  Meltzer, Harel, Koel, and Requicha}}]{maier03}
\bibinfo{author}{\bibfnamefont{S.}~\bibnamefont{Maier}},
  \bibinfo{author}{\bibfnamefont{P.}~\bibnamefont{Kik}},
  \bibinfo{author}{\bibfnamefont{H.}~\bibnamefont{Atwater}},
  \bibinfo{author}{\bibfnamefont{S.}~\bibnamefont{Meltzer}},
  \bibinfo{author}{\bibfnamefont{E.}~\bibnamefont{Harel}},
  \bibinfo{author}{\bibfnamefont{B.}~\bibnamefont{Koel}}, \bibnamefont{and}
  \bibinfo{author}{\bibfnamefont{A.}~\bibnamefont{Requicha}},
  \bibinfo{journal}{Nature Materials} \textbf{\bibinfo{volume}{2}},
  \bibinfo{pages}{229} (\bibinfo{year}{2003}).

\bibitem[{\citenamefont{Huter}(2004)}]{huter04}
\bibinfo{author}{\bibfnamefont{J.~H.} \bibnamefont{Huter},
  \bibfnamefont{E.~Fendler}}, \bibinfo{journal}{Adv. Mater.}
  \textbf{\bibinfo{volume}{16}}, \bibinfo{pages}{1685} (\bibinfo{year}{2004}).

\bibitem[{\citenamefont{Mizrahi et~al.}(1989)\citenamefont{Mizrahi, DeLong,
  Stegeman, Saifi, and Andrejco}}]{mizrahi89}
\bibinfo{author}{\bibfnamefont{V.}~\bibnamefont{Mizrahi}},
  \bibinfo{author}{\bibfnamefont{K.~W.} \bibnamefont{DeLong}},
  \bibinfo{author}{\bibfnamefont{G.~I.} \bibnamefont{Stegeman}},
  \bibinfo{author}{\bibfnamefont{M.~A.} \bibnamefont{Saifi}}, \bibnamefont{and}
  \bibinfo{author}{\bibfnamefont{M.~J.} \bibnamefont{Andrejco}},
  \bibinfo{journal}{Opt. Lett.} \textbf{\bibinfo{volume}{14}},
  \bibinfo{pages}{1140} (\bibinfo{year}{1989}).

\bibitem[{\citenamefont{Hartfield and Thompson}(1978)}]{hartfield78}
\bibinfo{author}{\bibfnamefont{E.}~\bibnamefont{Hartfield}} \bibnamefont{and}
  \bibinfo{author}{\bibfnamefont{B.}~\bibnamefont{Thompson}},
  \emph{\bibinfo{title}{Handbook of {O}ptics}}
  (\bibinfo{publisher}{McGraw-Hill}, \bibinfo{year}{1978}), chap.
  \bibinfo{chapter}{Optical modulators}, pp. \bibinfo{pages}{17--1 -- 17--24}.

\bibitem[{\citenamefont{Born and Wolf}(1980)}]{born80}
\bibinfo{author}{\bibfnamefont{M.}~\bibnamefont{Born}} \bibnamefont{and}
  \bibinfo{author}{\bibfnamefont{E.}~\bibnamefont{Wolf}},
  \emph{\bibinfo{title}{Principles of {O}ptics}} (\bibinfo{publisher}{Pergamon
  Press}, \bibinfo{year}{1980}), \bibinfo{edition}{6th} ed.

\bibitem[{\citenamefont{Sutherland}(1996)}]{sutherland96}
\bibinfo{author}{\bibfnamefont{R.~L.} \bibnamefont{Sutherland}},
  \emph{\bibinfo{title}{Handbook of {N}onlinear {O}ptics}}
  (\bibinfo{publisher}{Marcel Dekker, Inc}, \bibinfo{year}{1996}), pp.
  \bibinfo{pages}{295--384}.

\bibitem[{\citenamefont{Johnson et~al.}(1972)\citenamefont{Johnson, Christy,
  and W.}}]{johnson72}
\bibinfo{author}{\bibfnamefont{P.~B.} \bibnamefont{Johnson}},
  \bibinfo{author}{\bibnamefont{Christy}}, \bibnamefont{and}
  \bibinfo{author}{\bibfnamefont{R.}~\bibnamefont{W.}}, \bibinfo{journal}{Phys.
  Rev. B} \textbf{\bibinfo{volume}{6}}, \bibinfo{pages}{4370}
  (\bibinfo{year}{1972}).

\bibitem[{\citenamefont{Kang and Khurgin}(2006)}]{kang06}
\bibinfo{author}{\bibfnamefont{J.}~\bibnamefont{Kang}} \bibnamefont{and}
  \bibinfo{author}{\bibfnamefont{J.}~\bibnamefont{Khurgin}}
  (\bibinfo{organization}{Optical Society of America}, \bibinfo{address}{Long
  Beach, CA}, \bibinfo{year}{2006}), \bibinfo{number}{QTuI2},
  \bibinfo{note}{conference on lasers and electro-optics/quantum electronics
  and laser science and photonic applications, systems and technologies}.

\end{thebibliography}
\end{document}